\documentclass[aip,rsi,reprint,graphicx]{revtex4-1} 
\usepackage{color}
\usepackage{time,mathptmx,helvet}
\usepackage{amsmath}
\usepackage{graphicx}

\usepackage[
pdfstartview=XYZ,
bookmarks=false,
colorlinks=true,
linkcolor=blue,
urlcolor=blue,
citecolor=blue,
pdftex
]{hyperref}

\draft 

\begin{document}


\title{\large\textcolor{blue}{Absolute Calibration of a Time-Resolved High Resolution X-ray Spectrometer for the National Ignition Facility (invited)}} 

\thanks{Invited paper published as part of the Proceedings of the 22nd Topical Conference on 
High-Temperature Plasma Diagnostics, San Diego, California, April, 2018.\\}

\author{Lan~Gao}
\email{lgao@pppl.gov.\\}
\affiliation{Princeton Plasma Physics Laboratory, Princeton University, Princeton, NJ 08543, USA} 

\author{B. F. Kraus}
\affiliation{Princeton Plasma Physics Laboratory, Princeton University, Princeton, NJ 08543, USA}

\author{K. W. Hill}
\affiliation{Princeton Plasma Physics Laboratory, Princeton University, Princeton, NJ 08543, USA} 

\author{M. Bitter}
\affiliation{Princeton Plasma Physics Laboratory, Princeton University, Princeton, NJ 08543, USA} 

\author{P. Efthimion}
\affiliation{Princeton Plasma Physics Laboratory, Princeton University, Princeton, NJ 08543, USA} 

\author{M. B. Schneider}
\affiliation{Lawrence Livermore National Laboratory, Livermore, California 94550, USA} 

\author{A. G. MacPhee}
\affiliation{Lawrence Livermore National Laboratory, Livermore, California 94550, USA} 

\author{D. B. Thorn}
\affiliation{Lawrence Livermore National Laboratory, Livermore, California 94550, USA} 

\author{J. Kilkenny}
\affiliation{Lawrence Livermore National Laboratory, Livermore, California 94550, USA}

\author{J. Ayers}
\affiliation{Lawrence Livermore National Laboratory, Livermore, California 94550, USA}

\author{R. Kauffman}
\affiliation{Lawrence Livermore National Laboratory, Livermore, California 94550, USA}

\author{H. Chen}
\affiliation{Lawrence Livermore National Laboratory, Livermore, California 94550, USA}

\author{D. Nelson}
\affiliation{Laboratory for Laser Energetics, University of Rochester, Rochester, NY14623, USA} 

\date{\today}

\begin{abstract}

A high resolution, Diagnostic Instrument Manipulator (DIM)-based x-ray Bragg crystal spectrometer has been calibrated for and deployed at the National Ignition Facility (NIF) to diagnose plasma conditions in ignition capsules near stagnation times. The spectrometer has two conical crystals in the Hall geometry focusing rays from the Kr He$\alpha$, Ly$\alpha$, and He$\beta$ complexes onto a streak camera, with the physics objectives of measuring time-resolved electron density and temperature through observing Stark broadening and the relative intensities of dielectronic satellites. A third von H\'amos crystal that time-integrates the Kr He$\alpha$, He$\beta$ and intervening energy range provides in-situ calibration for the streak camera signals. The spectrometer has been absolutely calibrated using a microfocus x-ray source, an array of CCD and single-photon-counting detectors, and multiple K- and L-absorption edge filters at the Princeton Plasma Physics Laboratory (PPPL) x-ray laboratory. Measurements of the integrated reflectivity, energy range, and energy resolution for each crystal are discussed. These calibration data provide absolute x-ray signal levels for NIF measurements, enabling precise filter selection and comparisons to simulations. 

\end{abstract}

\maketitle 

\section{Introduction}

Precision measurements of the hot spot plasma conditions such as density and temperature for an implosion target at stagnation is of great importance for achieving ignition at the National Ignition Facility (NIF). \cite{lindl1995} Since the stagnated state is essentially evolving dramatically, direct measurements with time resolution of these parameters can provide more robust constraints on simulations that lead to ignition target designs. \cite{Lindl2004} Currently on the NIF, the ion temperature T$_{\rm i} $ is measured from neutron time-of-flight (NTOF) spectrometers \cite{Glebov_RSI_2010} and the density n$_{\rm e}$ is deduced from T$_{\rm i} $ assuming uniform plasma conditions for the hot spot. The NTOF data, besides being spatially and temporally integrated, is affected by plasma flows and implosion asymmetry resulting in ambiguity in the inference of T$_{\rm i} $. \cite{Spears_PoP_2015, Maria_PRE_2016}

High resolution x-ray spectroscopy is a well-established technique on the magnetic confinement fusion facilities that has successfully generated profiles of Doppler T$_{\rm i}$ , electron temperature T$_{\rm e}$  and the flow velocity. \cite{shi10,beiersdorfer10,hill10} It is therefore being adapted to the inertial confinement fusion and high energy density plasmas to provide detailed measurements on the plasma properties at extreme conditions. \cite{Nilson_RSI_2016} A Kr-doped gas symmetry capsule platform has been developed on the NIF to allow for x-ray spectroscopy of the Kr K-emission lines from implosion cores. \cite{Ma_RSI_2016} Kr He$\alpha$, He$\beta$, as well as satellite lines were successfully recorded with the NIF X-ray spectrometer (NXS)\cite{Perez_RSI_2014} when the capsule compressed and stagnated. The experiments demonstrated the feasibility of using a middle-z dopant in the capsule to extract x-ray spectroscopic information emitted from the hot spot. However, due to the low resolving power of NXS ( $\ge$ 60), it was not possible to infer n$_{\rm e}$ and T$_{\rm e}$ from the measured x-ray lines. \cite{Chen_PoP_2017}

A time-resolved, high resolution x-ray Bragg crystal spectrometer has been developed to measure the hot spot n$_{\rm e}$ and T$_{\rm e}$ on the NIF. \cite{Hill_IAEA_2016, Hill_RSI_2016} The instrument, named as dHIRES (DIM-based high resolution spectrometer), is a snout that mounts to a DIM Insertable Streak Camera (DISC)\cite{Kalantar_RSI_2001,Kimbrough_RSI_2001} and is fielded in a DIM for accurate positioning in the NIF chamber. The spectrometer has two conical crystals in the Hall geometry \cite{Hall_1984} focusing rays from the Kr He$\alpha$, Ly$\alpha$, and He$\beta$ complexes onto DISC with a temporal resolution of $\sim$30 ps. The physics goal is to measure time-resolved electron density and temperature through observing Stark broadening and the relative intensities of dielectronic satellites. A third von H\'amos crystal that time-integrates the entire energy range provides in-situ calibration for the streak camera signals. See Refs. [\onlinecite{Hill_IAEA_2016, Hill_RSI_2016}] for more details about the spectrometer design process, including feasibility of measuring n$_{\rm e}$ and T$_{\rm e}$ based on simulations, spectral resolution requirement, crystal selection, how to field three crystals into one cassette for DIM and avoid beam interference in the NIF chamber, etc.

Before dHIRES was deployed for NIF experiments, significant efforts had been made at the Princeton Plasma Physics Laboratory (PPPL) x-ray laboratory to fully calibrate the instrument. Absolute calibration of a spectrometer relates the detector signal level to the emission of a source. By measuring the throughput conversion factor between the source and detector signals, absolute brightness of the measured spectra can be known. Such practice is crucial for properly setting up the diagnostic based on pre-shot simulations, and post-shot comparisons between the experimental measurements and calculations from collisional-radiative codes.

This paper reports details of the calibration work for dHIRES. Sec.~\ref{sec:design} describes the final spectrometer design; Sec.~\ref{sec:calibration} discusses the calibration process including source alignment, crystal evaluation, energy calibration, source displacement error analysis, and the absolute throughput measurement for predicting NIF signal levels; and Sec.~\ref{sec:summary} provides summary and conclusions. 

\section{Spectrometer Design}
\label{sec:design}

\begin{figure}
\includegraphics[width=8.6 cm]{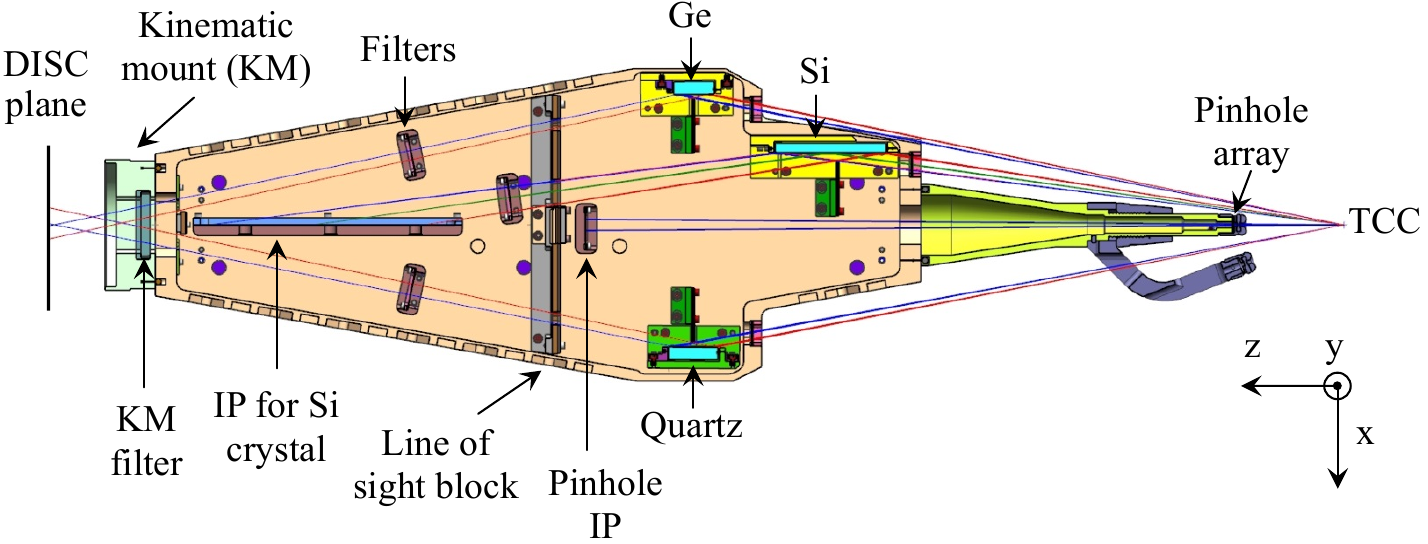}
\caption{\label{fig:spectrometer} dHIRES instrument layout. X-rays emitted from TCC and diffracted from each crystal are shown.}
\end{figure}

\begin{figure}[b]
\includegraphics[width=5 cm]{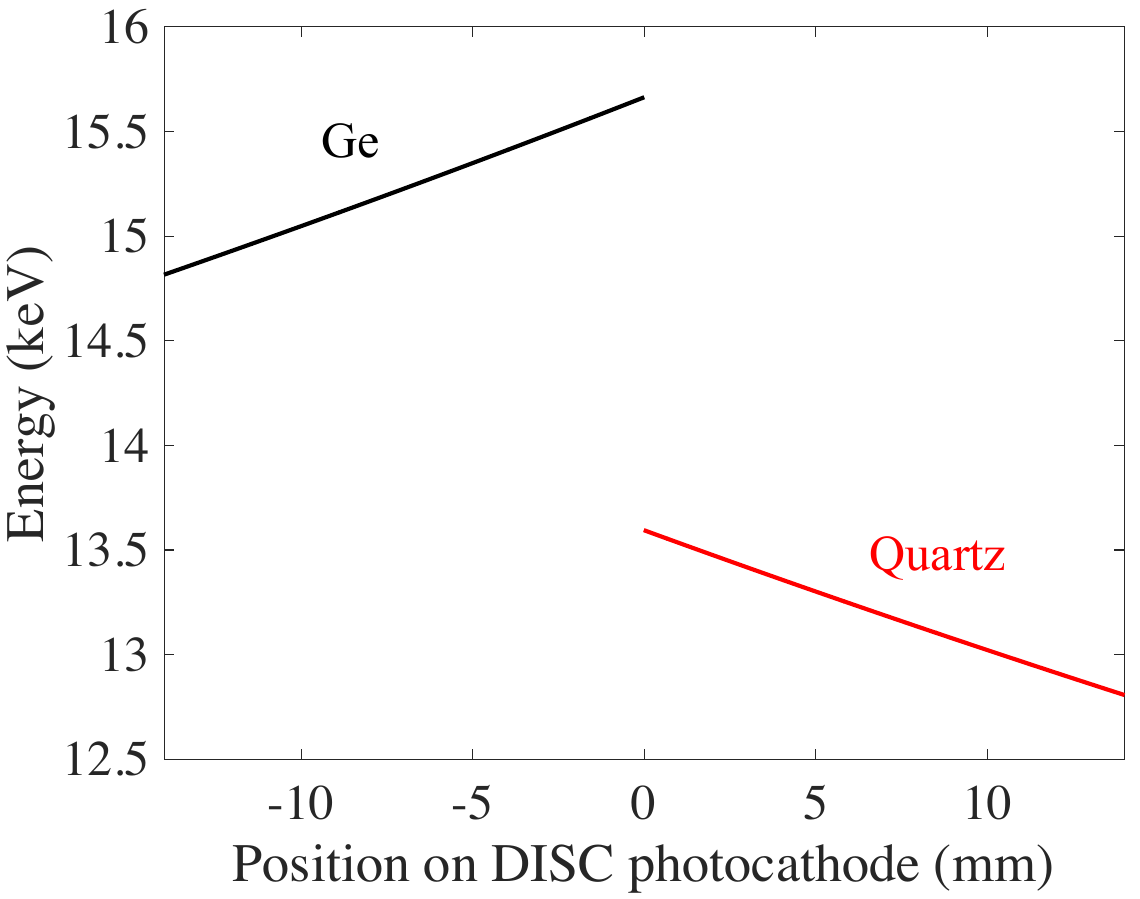}
\caption{\label{fig:SpecDispersion} Calculated spectral dispersion as a function of the position along the DISC photocathode for the Ge and quartz crystals.}
\end{figure}

The final design of dHIRES is based on three x-ray spectrometer channels and one imaging channel. Fig.~\ref{fig:spectrometer} shows the instrument layout for dHIRES. X-rays emitted from the target chamber center (TCC) generated by the compressing and stagnating plasmas in the hot spot pass through three entrance windows protected by blast shields, impinge onto three Bragg crystals, and then diffract onto their corresponding detectors. The conical Ge $\langle 220 \rangle $ crystal provides high resolution measurements of the Kr He$\beta$ lines, and the conical quartz $\langle110 \rangle $ crystal covers x-ray energies for Kr Ly$\alpha$ and He$\alpha$. Both crystals are in the Hall geometry \cite{Hall_1984} to provide a uniform sagittal focus along the DISC photocathode which is perpendicular to the spectrometer axis. For dHIRES, there is an additional rigid-body rotations for both crystals so that the Kr He$\beta$ and He$\alpha$ complexes are recorded on the lower and upper parts of the DISC photocathode respectively resulting in time evolution of the spectra. The crystal choices are made by optimizing the resolving power for each channel and positioning both spectra on the DISC photocathode with an effective length of 24 mm without overlapping the energies of interest. Distance from the DISC plane to TCC is 1160 mm. Since the Ge and quartz crystals have different reflectivities and the DISC camera is not calibrated, a third von H\'amos spectrometer using a cylindrical Si $\langle 111 \rangle $ crystal is used to diffract the entire energy range of our interest onto an image plate (IP). The IP is located slightly below and almost parallel to the dHIRES central axis. This measurement is used to normalize the two time-resolved channels. The imaging channel relies on an array of four pinholes that are located by the instrument snout near the TCC. Broadband x-rays propagating through the pinholes are recorded by another IP at the center of the dHIRES cassette. By using differentiated filters for the pinhole images, 2D time-integrated T$_{\rm e} $ from the emitting core can be measured. \cite{Ma_RSI_2012}

Fig.~\ref{fig:SpecDispersion} shows the calculated spectral dispersion relative to the length of the DISC photocathode. The distances are referenced to the center of the DISC photocathode on the DIM axis. Therefore x-ray energy increases toward the center for both crystals on the DISC camera. Given that the photocathode has an effective length of 24 mm and there is $\sim$1 mm gap between the two spectra, the energy range covered by the Ge crystal is 14.9 keV to 15.6 keV, and it is 12.8 keV to 13.6 keV for the quartz.

\section{dHIRES Calibration}
\label{sec:calibration}

\begin{figure}[b]
\includegraphics[width=8 cm]{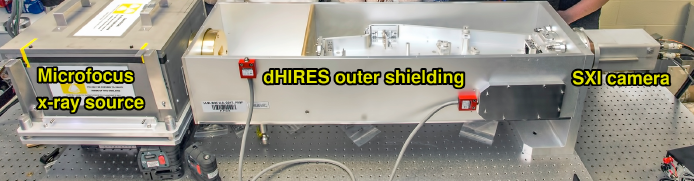}
\caption{\label{fig:setup} Setup for dHIRES calibration at the PPPL x-ray laboratory. The lid for the aluminum box is removed and the dHIRES cassette can be seen.}
\end{figure}

The dHIRES cassette was manufactured at the Laboratory for Laser Energetics, University of Rochester. The crystals were shaped and mounted by Inrad Optics. The whole spectrometer was assembled, aligned, and fully calibrated in the PPPL X-ray Laboratory. Final setup of the calibration system is shown in Fig.~\ref{fig:setup}.

A point-like x-ray source was generated by a Hamamatsu microfocus x-ray tube with a tungsten anode. The source size measured with knife edges is $\sim$8.4 $\mu$m. \cite{Lu14} The x-ray source box is supported by a plate that has a pair of bolt and screw on each corner. By pressing or releasing the screws, it is possible to change the source height, rotation and tilt. The source box and the plate sit on top of a linear translation stage to provide 2D source adjustments, so that the distances between the source and crystals and detector can be varied. 
 
The dHIRES cassette was mounted inside an aluminum box to protect personnel from radiation leakage. A removable lid enables x-ray exposures with the lid in place, as well as convenient adjustments for spectrometer alignment and calibration. Aluminum, copper and stainless steel shieldings were also used to ensure the entire calibration system was safe for operation.

A platform was built using aluminum I-beams and box beams and plates to raise the central axis of the dHIRES cassette to the level of the x-ray beam. The platform was bolted securely to the optical table, and the dHIRES cassette was firmly attached to the platform to prevent movement of the spectrometer during calibration.

The majority of calibration exposures were detected by a SXI SI-800 x-ray CCD with 2084 $\times$ 2084 pixels with a pixel size 24 $\times$ 24 $\mu$m. The active area is large enough to simultaneously cover the spectra from Ge and quartz crystals. The detector rests on an extension bracket at the back end of the dHIRES cassette.

In the following section, details of the calibration work are reported. After the x-ray source was aligned, we evaluated the crystal performance and positioned the crystals so that the spectra from Ge and quartz were in line with each other. We then practiced the source displacement and DIM insertion error analysis to understand their effect on the spectra. Tremendous work was carried out to measure the absolute throughputs for all three crystals which enable calibration of the measured spectra for NIF experiments.

\subsection{Source and Crystal Alignment}
\begin{figure}
\includegraphics[width=7 cm]{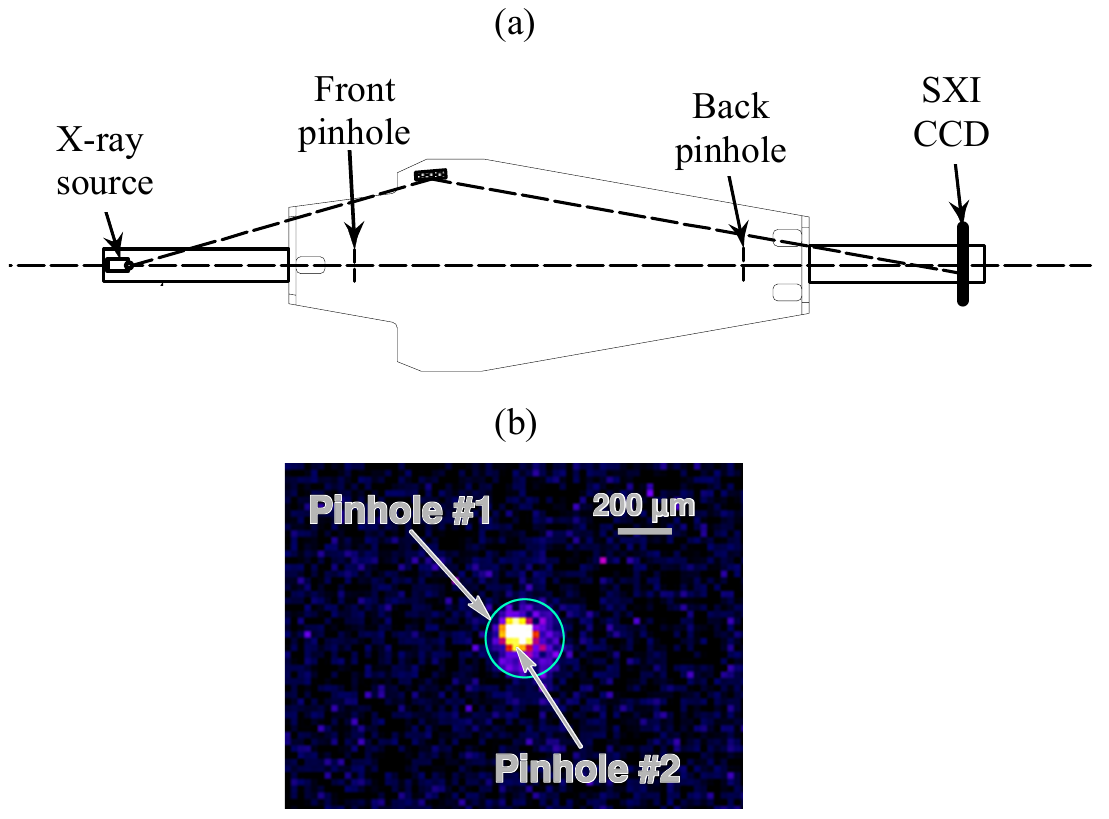}
\caption{\label{fig:sourcealign1} (a) A schematic showing how the x-ray source is aligned to the dHIRES/DIM axis determined by a pair of front and back pinholes. (b) is an example x-ray image where two 100-$\mu$m pinholes were aligned to each other within one pixel.}
\end{figure}

The first step of the calibration work was to align the x-ray source to the dHIRES central axis which is also the DIM axis. As is shown in Fig.~\ref{fig:sourcealign1}(a), there is one pinhole in the front and one pinhole in the back of dHIRES, and the line connecting the two pinholes defines its axis. By tuning the source height, rotation and tilt angle through the translation stage and the screws on the source support plate, x-rays emitted at the source propagated through both pinholes and formed a spot on the CCD. We started with 500-$\mu$m diameter pinholes and graduated to smaller diameter pinholes to refine the alignment. Fig.~\ref{fig:sourcealign1}(b) presents an example x-ray image where two 100-$\mu$m diameter pinholes were used. The pinhole image $\#$1 with a larger spot diameter represents the front pinhole due to larger magnification compared to the pinhole image $\#$2. The two images were aligned to each other within one pixel (24 $\mu$m), indicating the source was aligned with the dHIRES axis. Following the dHIRES axis, the source was then put at the theoretical location where the TCC should be. This was verified by measuring the distance from the source to the SXI CCD detector plane to be 1160 mm.

\begin{figure}
\includegraphics[width=7 cm]{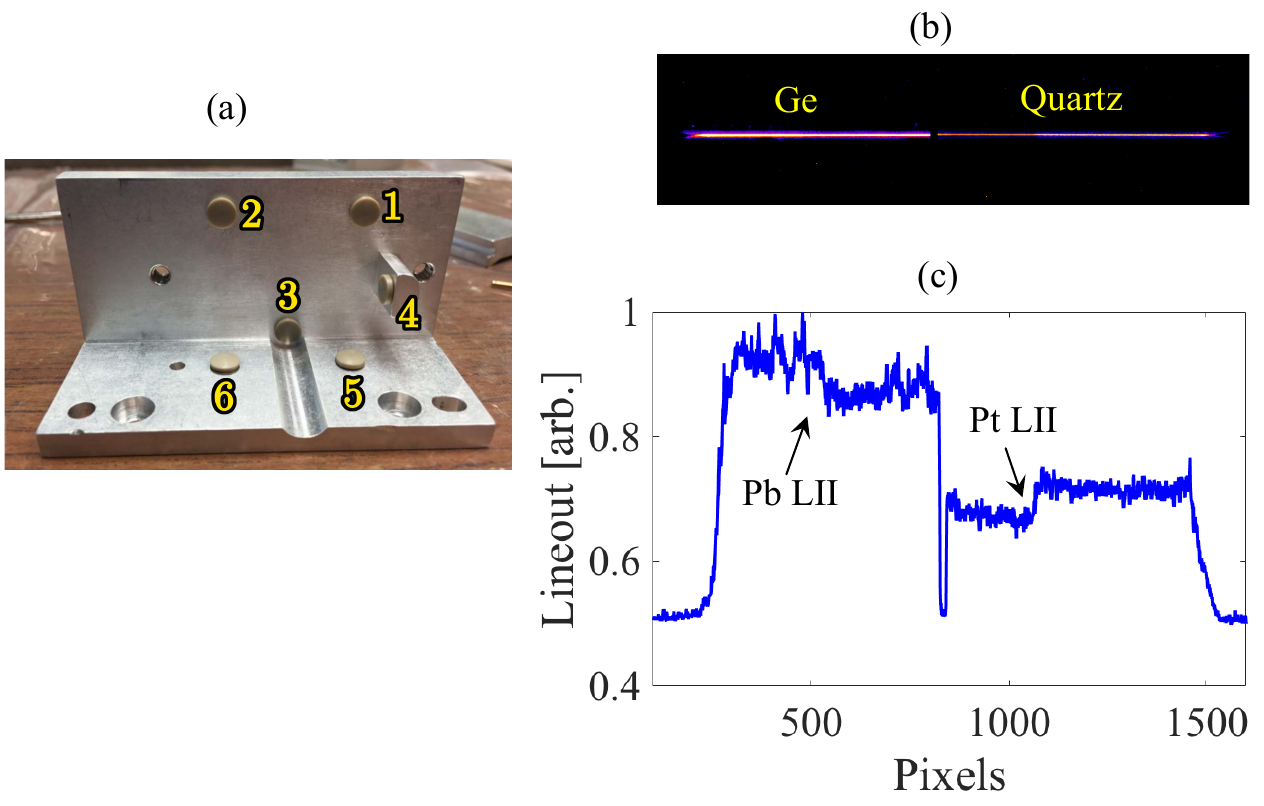}
\caption{\label{fig:sourcealign2} (a) six rubber pads on the crystal holder to provide six degree of freedom for crystal adjustment, (b) two crystal foci at the DISC plane after final adjustment of the crystals, and (c) horizontal lineout of the spectra shown in (b). }
\end{figure}

The crystals were precisely positioned by adjusting the crystal mounts. As is shown in Fig.~\ref{fig:sourcealign2}(a), there are six exchangeable rubber pads on the crystal holder. The manufactured pads have various thicknesses in 24 $\mu$m increments. 12 $\mu$m-thick shims were also made to put under the pads for finer adjustments when necessary. By choosing the corresponding pad/shim thicknesses, it is possible to achieve six degrees of freedom to change crystal coordinates and orientation. For example, pads 1, 3 and 2 with differentially increasing or reducing thicknesses will result in crystal rotation, while equal thickness change for them will move the crystal in x direction. By iteratively adjusting the crystal mounts, spectra from the Ge and quartz crystals were perfectly in line with each other, as is shown in Fig.~\ref{fig:sourcealign2}(b). Line spectrum on the left is from the Ge crystal and the right one is from quartz. The quartz crystal focus is less bright due to lower integrated reflectivity than the Ge crystal. Both spectra were aligned to the center of the pinhole image. This means that in actual NIF experiments, the line foci would enter the DISC slit and get streaked. Fig.~\ref{fig:sourcealign2}(c) is a horizontal lineout of the line spectra. Signal drop for Ge and quartz crystals are due to Pb and Pt LII edges which we used for energy calibration and will be discussed in Sec.~\ref{sec:dispersion}. There is signal gap in between the two foci. This is caused by a Ta knife edge at the back of the spectrometer (see Sec.~\ref{sec:shift}).

\begin{figure}[b]
\includegraphics[width=8 cm]{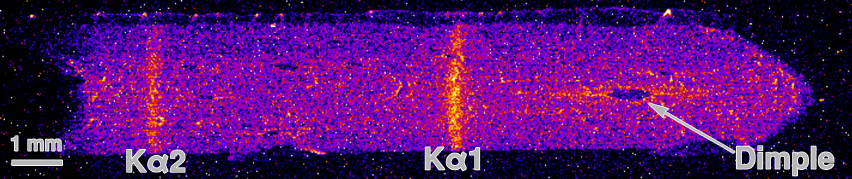}
\caption{\label{fig:defects} A defocused image from the Ge conical crystal. Crystal surface structures as well as W K$\alpha$ lines in 4$^\text{th}$ order are viewed.}
\end{figure}

\subsection{Crystal Evaluation}

The required crystal quality for all dHIRES crystals was determined based on high resolving power and sufficient signal level. The actual crystals received from the manufacturer were evaluated for comparison with theoretical calculations of ideal crystals. Fig.~\ref{fig:defects} shows a defocused spectral image from the Ge crystal, rotated out of alignment by $\sim$ 0.5$^{\circ}$. Rotating the crystal out of alignment has two effects. Firstly, it allows access to new x-ray energies. W K$\alpha$1 and K$\alpha$2 lines in 4$^\text{th}$ order were observed in this new configuration. Analysis of the line width compared to the natural width values led an estimated resolving power of $\sim$2000 for the Ge crystal. Secondly, it generates a defocused image where crystal surface structures are mapped. In Fig.~\ref{fig:defects}, one relatively large dimple was observed due to crystal detachment from the substrate. By running this configuration for more than 20 hours to accumulate photon statistics, the measured signal drop due to the observed dimple was $<$10$\%$. The measured crystal defect was included in the instrument function for calibrating the crystal throughput. 

\subsection{Energy Calibration and Wavelength Dispersion}\label{sec:dispersion}

\begin{figure}
\includegraphics[width=8 cm]{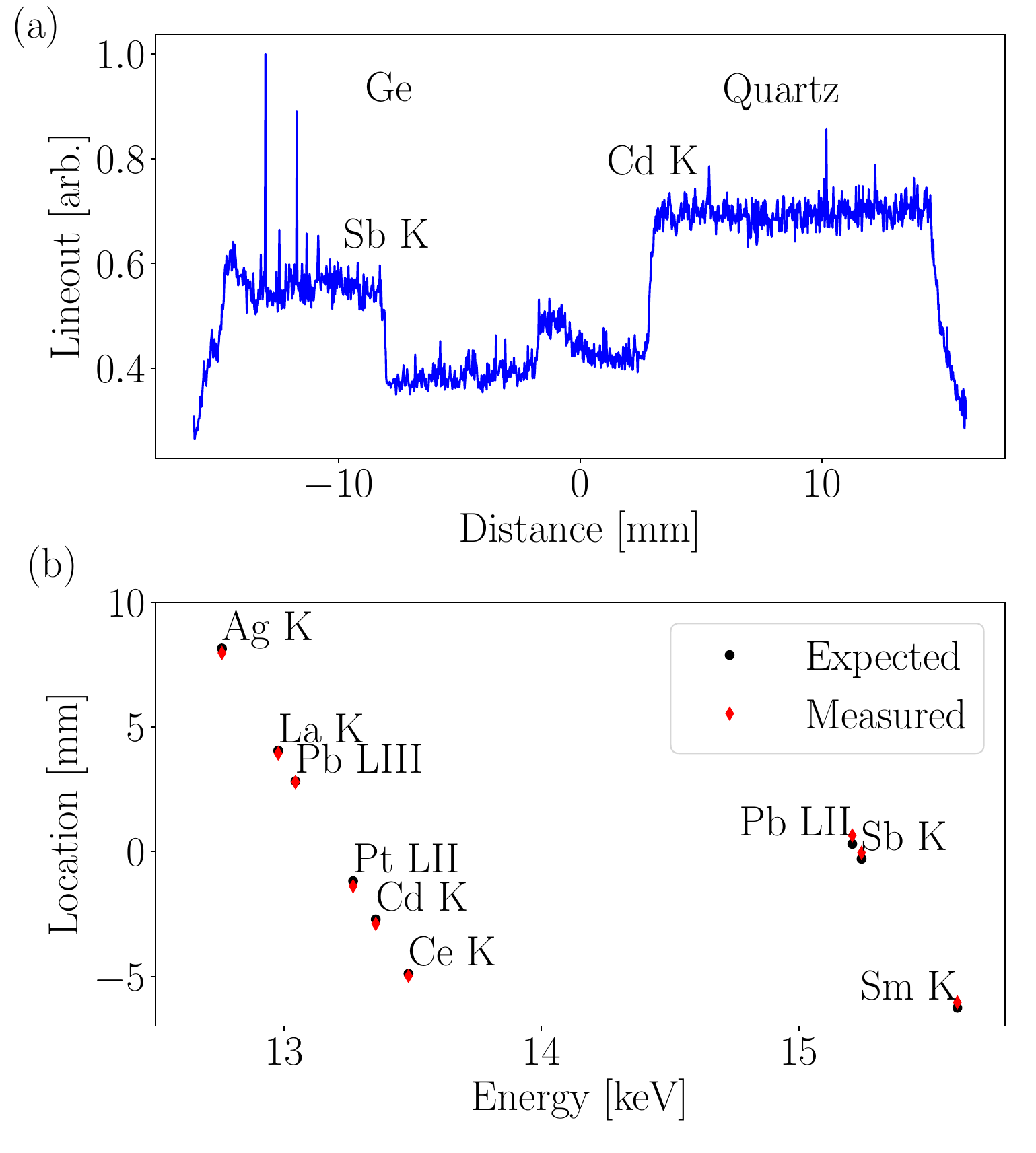}
\caption{\label{fig:dispersion} (a) The Sb and Cd K edges in 2$^\text{nd}$ order were observed in the spectral lineout for Ge and quartz respectively. (b) The measured location of absorption edges is consistent with the calculated geometrical dispersion.}
\end{figure}

The energy calibration was determined by observing the abrupt decrease in continuum intensity as a function of energy at well-known K and L absorption edges.\cite{Ingalls_JAP_1980,Bhalerao_JAP_2010} Within the energy range for each crystal (14.9 - 15.6 keV for Ge and 12.8 - 13.6 keV for quartz), a list of elements were identified. Fig.~\ref{fig:dispersion}(a) shows that sharp K edges in 2$^\text{nd}$ order for Sb and Cd were identified within 1- 2 pixels in the spectral lineout for Ge and quartz respectively. The Sb K edge in 2$^\text{nd}$ order has an energy of 15.246 keV, and the Cd K edge in 2$^\text{nd}$ order has an energy of 13.356 keV. This indicates that accurate energy calibration can be achieved using this method. Fig.~\ref{fig:dispersion}(b) presents the location of the measured absorption edges for all selected elements for both crystals. The experimentally measured dispersion is in good agreement with the calculated geometrical dispersion as is shown in Fig.~\ref{fig:SpecDispersion}, confirming the alignment of dHIRES in the lab setting.

\subsection{Spectrum Manipulation}\label{sec:shift}

\begin{figure}
\includegraphics[width=4.5 cm]{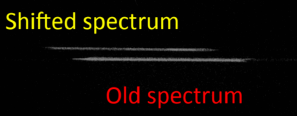}
\caption{\label{fig:shift} Spectrum shift from rigid-body rotation of the crystal.}
\end{figure}

When designing dHIRES, the two conical crystals were rotated about the x-ray source by rigid-body rotation to avoid significant overlap between the two spectra from Ge and quartz in the focal plane. This procedure shifts the spectrum spatially while maintaining their sagittal focus. As a result, the Kr He$\beta$ and He$\alpha$ complexes diffracted from Ge and quartz crystals are recorded on the lower and upper parts of the DISC photocathode respectively. This design was further tested in the lab. By using the adjustable mounting pads in each crystal holder, the necessary repositioning of each crystal for rigid-body rotation could be calculated exactly. Firstly, the necessary rotation angle for a desired energy shift was calculated. Following the coordinate system transformation matrix, the new coordinates of the crystal mounts were calculated. We then used shims and rubber pads to position the crystal at the new location. Fig.~\ref{fig:shift} shows a shifted spectrum consistent with our calculations and with the sagittal focus maintained, confirming the dHIRES design and construction.

To ensure that DISC captures all x-rays of the energies of our interest, the crystals were made larger than necessary, leading to signal overlap at the center between Ge and quartz. For example, in Fig.~\ref{fig:dispersion}(a), signal rise in the center is seen due to signal overlapping. Ta knife edges at the back of dHIRES and in front of DISC cut off the high-energy part of the spectrum to prevent such overlap, as is shown in Fig.~\ref{fig:sourcealign2}(b).

In addition, we have also calibrated how to adjust the spectrum height, tilt, and rotation using different adjustment pads. A calibration table was made to achieve spectrum adjustments offsite at NIF when needed.

\subsection{Source Displacement and DIM Insertion Error}

\begin{figure*}
\includegraphics[width=16 cm]{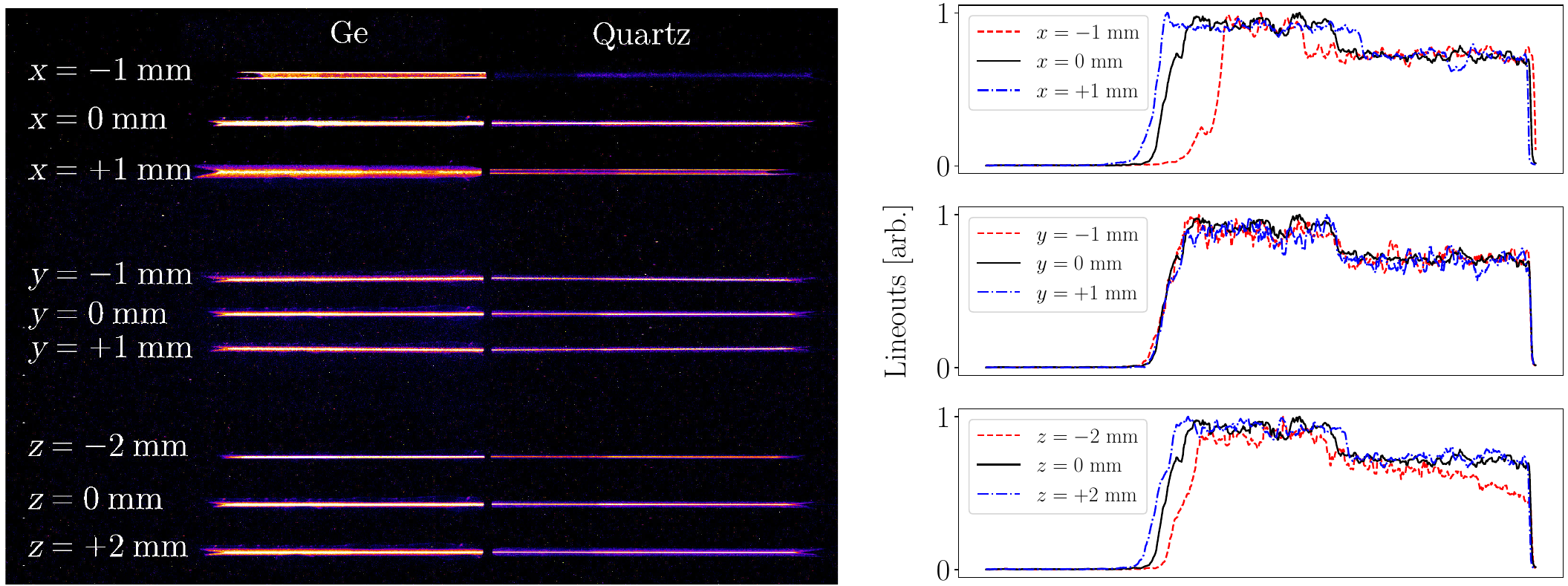}
\caption{\label{fig:insertion} Left panel: spectral change as a result of source displacement and DIM insertion error. Right panel: corresponding horizontal lineouts for the line foci for Ge.}
\end{figure*}

The aforementioned work was all conducted with the source on the DIM axis and at the theoretical location TCC. In actual NIF experiments, there are source positioning and DIM insertion errors that degrade the spectrometer performance. It is therefore of great importance to calibrate the effect of these errors on the spectra. Up to now, the lateral source positioning error at NIF is $\le$ $\pm$250 $\mu$m and the DIM insertion error is $\le$ $\pm$2 mm. The effects of these errors on crystal foci were quantified in the lab by positioning the source at the offset locations using the translation stage and the up-and-down screws. After adjustments, the focal properties of each crystal were monitored. Here $x$ is the lateral offset, $y$ is the offset vertically out of the cassette plane, and z is the DIM insertion error along the instrument axis, as are noted in Fig.~\ref{fig:spectrometer}.

The DIM insertion error induces an energy shift. As is shown in Fig.~\ref{fig:insertion} left panel, as the source was gradually moved from -2 mm to +2 mm, spectra moved out. This can be easily understood because the Bragg angle becomes larger when the source moves toward dHIRES and the energy range covered by the crystal therefore shifts to the lower energy side. Horizontal lineouts of the line foci for Ge are shown in the right panel where signal drops due to Pb LII edge are observed. While the entire spectrum moved out, the Pb LII edge at 15.207 keV moved in. The measured $\Delta E/\Delta z$ is about 10 eV per mm by monitoring this edge shift.

Variations in $y$ displace foci vertically without affecting energy. When the source was placed 1 mm below the theoretical height, the spectra became 1 mm above the theoretical center line. Similar results were seen when the source was placed in the other direction. 

Lateral variations in $x$ defocus spectra and shift energies. The experimental results are shown in the top row of Fig.~\ref{fig:insertion} left panel. When the source is placed at $x =$ +1 mm (closer to the quartz crystal and further from the Ge crystal), both spectra are shifted, with Ge covering lower energies and quartz covering higher energies. The measured $\Delta E/\Delta x$ is about 6.5 eV per 100 $\mu$m by monitoring the Pb LII edge shift. In addition, the focal shape is changed dramatically. While the fish-bone like shape maintains for Ge, the central part of the spectrum is less bright and blurring. Two horizontal peaks are observed for quartz. All of these features are consistent with our ray-tracing model calculations.

In actual NIF experiments, the measured energy shift and shape change in the spectra would be due to a combination of the source displacement and DIM insertion errors. It is possible to extract these fielding errors by quantifying the spectral change and comparing with our calibrations.

\subsection{Absolute throughput measurement and NIF Signal Level Prediction}

The majority part of the calibration work for dHIRES was measurement of the throughput for all three crystals which is essential for calibrating the spectra signal level. It enables inference of integrated reflectivities from each crystal, proper filtering of the detector signal based on theoretical predictions, and post-shot comparisons between measured spectral brightness and calculations from collisional-radiative codes.

The procedure for absolutely calibrating each crystal in dHIRES is as follows. First, a single-photon-counting detector measured the spectral intensity of the Hamamatsu microfocus x-ray source. Second, the same detector was used to measure 1-mm segments of the diffracted ray focus from each crystal, giving a quantity of photons per unit length. Third, a second detector with 2D imaging capability was used to measure the same foci, thus checking the variation in crystal reflectivity from low to high energies. Fourth, the measured intensities were adjusted where necessary to correct for detector quantum efficiency. Finally, the diffracted ray intensity $\left[ \text{J}/\text{mm} \right]$ is divided by the source spectral intensity $\left[ \text{J}/\text{sr keV} \right]$ to yield the throughput conversion factor $\left[ \text{sr keV}/\text{mm} \right]$.

Two single-photon-counting detectors were used for the outlined measurements. For spatially-integrated spectra, most measurements used an Amptek X-123 CdTe detector. The 1-mm-thick, 17-mm$^2$ solid state active area ensured nearly 100\% quantum efficiency for x-ray energies of interest. An integrated pulse height analyzer can achieve spectral resolution of $\sim$500 eV, large compared to the diffracted crystal rays but small enough to adequately measure continuum emission from the x-ray source. A second detector, the Dectris Pilatus 100K-S, can spatially resolve x-ray signals over an active area of 84 mm $\times$ 34 mm with a pixel size of 172 $\mu$m. The Pilatus quantum efficiency varies from 75\% to 50\% from 13 to 16 keV due to a 320 $\mu$m-thick Si sensor. Each of these detectors was mounted at the DISC plane in place of the SXI CCD camera inside a custom-made aluminum and tungsten shielding box with access for electronic wires and fan cooling.

\begin{figure}
\includegraphics[width=8 cm]{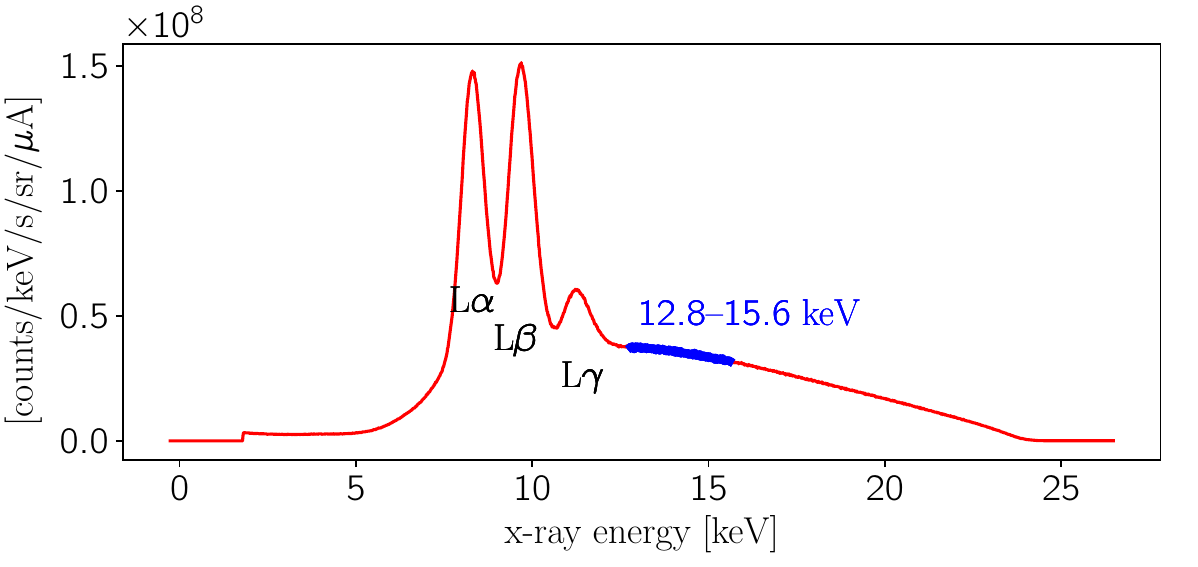}
\caption{\label{fig:source} The tungsten anode x-ray source emission spectrum, measured by the Amptek CdTe detector.}
\end{figure}

The emission of the microfocus source was recorded with the Amptek detector by placing the active area directly within the unfiltered x-ray beam. For this measurement and all others here, the microfocus tube was operated with a tungsten anode held at a fixed voltage of 24 kV. The intensity of the measured signal as a function of x-ray energy is established by this anode voltage, up to a proportionality constant that depends only on the current of the x-ray source (0--200 $\mu$A) and the exposure time. Emission from the source is steady in time as independently measured during all exposures by a second Amptek detector operated in multi-channel scaler mode. Detectors operated this manner show total counts in a region of the spectrum per one second interval as a function of time. This time history of energy-integrated source x-ray signal exhibits bin-to-bin variation of about 2\%, consistent with statistical error $1/\sqrt{N}$ for $N = 2000$ photons/s.

By optimizing the detector settings to eliminate electronic noise and avoid spectral pile-up, and accounting for the path difference when the source x-rays are not diffracted by the crystals and directly measured by the Amptek detector, the source spectrum was measured as shown in Fig.~\ref{fig:source}. The spectrum is attenuated on the low energy end by transmission through air, and falls to zero emission at $E = 24$ keV due to the anode voltage. Three unresolved tungsten complexes are visible, L$\alpha$, L$\beta$, and L$\gamma$, but none of these lines overlap with the continuum emission relevant to dHIRES from 12--16 keV. In this energy interval, the source emission $I_0$ is shown to vary with energy such that $I_0 = 3$--$5 \times 10^7$ photons/keV/sr/s/$\mu$A.

The next step in the calibration involves measurement of the diffracted x-ray intensity from each crystal, resolved as a function of length across the focus. Since the Amptek detector spatially integrates all photons that interact with its active area, it had to be modified to restrict access to a narrow band. To this end, the detector was fitted with a precision 1-mm-wide brass slit, aligned perpendicularly to the focus to allow an interval of approximate energy width 50 eV to focus on the detector. The detector was positioned at an 11$^{\circ}$ inclination to the DIM axis, so that rays diffracted from the crystal could pass unimpededly through parallel edges of the slit. Thus, the signal measured is directly in units of photons/mm, and is directly analogous to photons/pixel measured on detectors at NIF.

\begin{figure}
\includegraphics[width=0.45\textwidth]{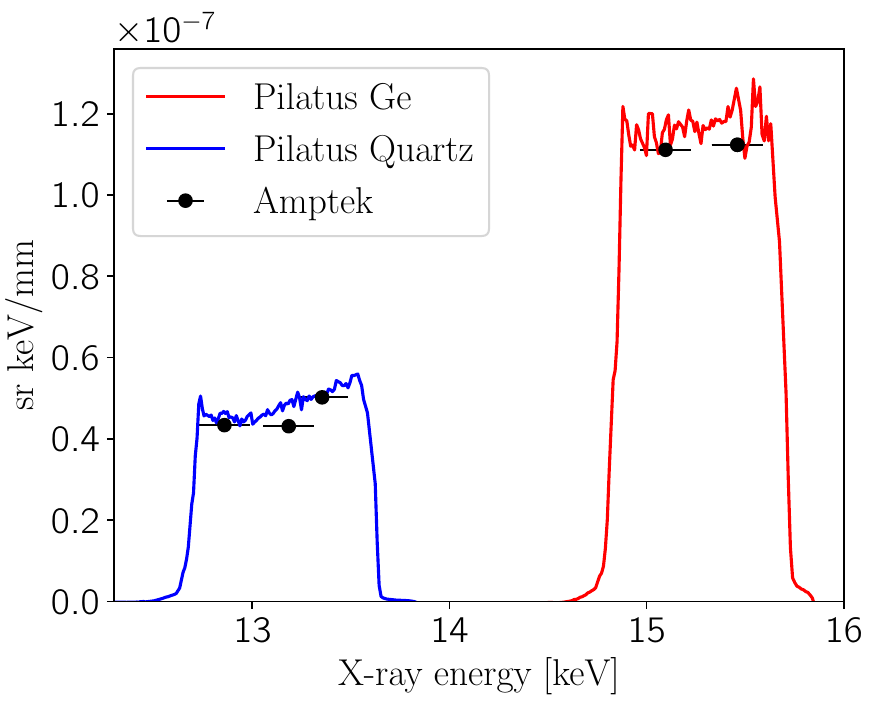}
\caption{\label{fig:throughput} The throughput conversion factor for both quartz and Ge crystals, calculated from the diffracted ray intensity divided by the source spectral intensity. }
\end{figure}

Such diffracted ray measurements were made at several locations and with high anode current (200 $\mu$A) to minimize exposure time. For each measurement, enough count rate was ensured so that the statistical error is $<$ 1\%. At each location, the detector measured a spectrum of diffracted rays. Even though these spectra had artificially high widths $>500$ eV due to the poor spectral resolution of the Amptek detector, the peak energy of the spectra shifted as expected from low to high energy as the detector was moved further from the source. The spectrally-integrated counts from each of these locations provided a photon count per mm along the focus at three distinct locations along the quartz crystal focus, giving values of roughly 2.0 photons/mm/s/$\mu$A, and two along that of Ge, with values around 3.8 photons/mm/s/$\mu$A. These values were corroborated with the 2D Pilatus detector, which was able to image the entire focus of each crystal at once. Once corrected for quantum efficiency, the lineouts of each Pilatus image are representative of the reflectivity variations across each crystal, including dips from small defects.

Finally, the throughput conversion factor can be derived by dividing the diffracted x-ray intensity by the source intensity, given that the energies of each spatial location are well known from our absorption edge dispersion analysis (see Sec.~\ref{sec:dispersion}). The result is shown for both Amptek and Pilatus detectors in Fig.~\ref{fig:throughput}. The independent measurements from each detector agree well. It can also be seen that small variations in reflectivity of each crystal are dwarfed by the difference between the two crystals. The factor for quartz, $5 \times 10^{-8}$ sr keV/mm, is less than half of the factor for Ge, $1.2 \times 10^{-7}$ sr keV/mm.

These conversion factors can be translated into integrated reflectivities and ultimately diffraction Darwin widths\cite{Stepanov} by making several assumptions. The mean wavelength dispersion of the quartz (Ge) crystal was measured to be 56 (58) eV/mm. Dividing the throughput conversion factors by these mean wavelength dispersions gives the effective solid angle from the source of each crystal: 0.90 $\mu$sr for quartz and 2.1 $\mu$sr for Ge. Each crystal is $\sim$59 cm from the source and 2.8 cm in height, so that the integrated reflectivities are 19 $\mu$rad for quartz and 44 $\mu$rad for Ge. These numbers are reasonably consistent with integrated reflectivities calculated for perfect thin \cite{Stepanov} and bent crystals \cite{SanchezdelRio}, with measured values about two times higher than predictions. If the peak reflectivity of the crystals is assumed to be near unity, the integrated reflectivity values are approximately equal to the rocking curve widths for each crystal. Based on these integrated reflectivity measurements, the energy resolution for quartz and Ge is 1.6 eV and 3.1eV respectively.

\section{Summary and Conclusions}
\label{sec:summary}

\begin{table}[b]
\vspace{0.1cm}
\tabcolsep=0.03cm
\begin{tabular}{c| c |c |c |c} 
\hline
\hline 
& Units & Kr He He$\alpha$ & Kr He$\beta$ & Full range \\
\hline 
\hline 
Crystal & &quartz $\langle110 \rangle $ & Ge $\langle 220 \rangle $  & Si $\langle 111 \rangle $ \\
Energy range & keV & 12.8 - 13.6 & 14.9 - 15.6 & 12.8 - 15.6 \\
Geometry & & Hall Conical & Hall Conical & Cylindrical \\
Detector & & DISC & DISC & IP \\
Rocking curve $\Delta\theta_{\rm flat}$ & $\mu$rad & 7.2 & 30 & 19.3 \\
$\Delta E_{\rm flat} \rm (theory) $ &eV & 0.49 & 2.2 & 2.0 \\
$\Delta E_{\rm curved} \rm (measurement) $ & eV&1.6&3.1&5.1 \\
$\Delta E$ (100 $\mu$m source)& eV & 5.6 & 6.3 & 10.9 \\
$\Delta E$ (100 $\mu$m detector)& eV &5.5 & 6.0 & 1.5\\
$\Delta E_{\rm total}$ & eV & 8.0 & 9.2 & 12.1 \\
$E/\Delta E $ (estimated) & & 1650 & 1600 & 1170 \\
\hline 
\hline 
\end{tabular}
\caption{Overview of the dHIRES parameters.}  
\label{tab:parameter} 
\end{table}

In summary, a time-resolved, high resolution x-ray spectrometer dHIRES that measures Kr He-like K lines for the NIF has been developed and fully calibrated at the PPPL x-ray laboratory. A point-like x-ray source was aligned to the spectrometer axis using a pair of pinholes, followed by accurate crystal positioning to align the spectra on the CCD camera. This ensures that the spectra from the two conical crystals will fall onto the DISC photocathode for time sweeping during actual NIF experiments. The crystal quality was evaluated by rotating the crystal to defocus the spectrum and map the surface structures. A series of K and L absorption edges were used to calibrate energy and measure the spectral dispersion on the detector. Spectrum adjustment was tested using the method of rigid-body rotation and knife edges. Source displacement and DIM insertion error which occur in actual experiments were tested to understand their effect on the spectra. The integrated reflectivity and throughput conversion factor were measured for all three crystals allowing calibration of the signal brightness for the measured spectra.

Table~\ref{tab:parameter} lists an overview of the dHIRES parameters. These measurements show that the broadening due to penetration into the crystal is smaller than XOP predictions, \cite{Hill_IAEA_2016} and the spectral resolution is not limited by diffraction width. Broadening due to source size and detector spatial resolution for the actual experiments can be dominant. 

Up till now, dHIRES has been fielded in five NIF shots and successfully recorded high quality x-ray spectroscopic data. Data analysis is underway and will be reported in a subsequent paper.

\begin{acknowledgments}
This work was performed under the auspices of the U.S. Department of Energy by Princeton Plasma Physics Laboratory under contract DE-AC02-09CH11466 and by Lawrence Livermore National Laboratory under contract DE-AC52-07NA27344.
\end{acknowledgments}

\bibliography{X-ray}

@article{Perez_RSI_2014,
author = {P\'erez,F.  and Kemp,G. E.  and Regan,S. P.  and Barrios,M. A.  and Pino,J.  and Scott,H.  and Ayers,S.  and Chen,H.  and Emig,J.  and Colvin,J. D.  and Bedzyk,M.  and Shoup,M. J.  and Agliata,A.  and Yaakobi,B.  and Marshall,F. J.  and Hamilton,R. A.  and Jaquez,J.  and Farrell,M.  and Nikroo,A.  and Fournier,K. B. },
title = {The NIF x-ray spectrometer calibration campaign at Omega},
journal = {Review of Scientific Instruments},
volume = {85},
number = {11},
pages = {11D613},
year = {2014},
doi = {10.1063/1.4891054},
}

@article{Kalantar_RSI_2001,
author = {D. H. Kalantar and P. M. Bell and T. S. Perry and N. Sewall and J. Kimbrough and F. Weber and C. Diamond and K. Piston},
title = {Optimizing data recording for the NIF core diagnostic x-ray streak camera},
journal = {Review of Scientific Instruments},
volume = {72},
number = {1},
pages = {751-754},
year = {2001},
doi = {10.1063/1.1318263},
}

@article{Kimbrough_RSI_2001,
author = {J. R. Kimbrough and P. M. Bell and G. B. Christianson and F. D. Lee and D. H. Kalantar and T. S. Perry and N. R. Sewall and A. J. Wootton},
title = {National Ignition Facility core x-ray streak camera},
journal = {Review of Scientific Instruments},
volume = {72},
number = {1},
pages = {748-750},
year = {2001},
doi = {10.1063/1.1318262},
}

@article{Nilson_RSI_2016,
author = {P. M. Nilson and F. Ehrne and C. Mileham and D. Mastrosimone and R. K. Jungquist and C. Taylor and C. R. Stillman and S. T. Ivancic and R. Boni and J. Hassett and D. J. Lonobile and R. W. Kidder and M. J. {Shoup III} and A. A. Solodov and C. Stoeckl and W. Theobald and D. H. Froula and K. W. Hill and L. Gao and M. Bitter and P. Efthimion and D. D. Meyerhofer},
title = {A high-resolving-power x-ray spectrometer for the OMEGA EP Laser (invited)},
journal = {Review of Scientific Instruments},
volume = {87},
number = {11},
pages = {11D504},
year = {2016},
doi = {10.1063/1.4961076},
}

@article{Chen_PoP_2017,
author = {Hui Chen and T. Ma and R. Nora and M. A. Barrios and H. A. Scott and M. B. Schneider and L. Berzak Hopkins and D. T. Casey and B. A. Hammel and L. C. Jarrott and O. L. Landen and P. K. Patel and M. J. Rosenberg and B. K. Spears},
title = {On krypton-doped capsule implosion experiments at the National Ignition Facility},
journal = {Physics of Plasmas},
volume = {24},
number = {7},
pages = {072715},
year = {2017},
doi = {10.1063/1.4993049},
}

@article{Ma_RSI_2016,
author = {T. Ma and H. Chen and P. K. Patel and M. B. Schneider and M. A. Barrios and D. T. Casey and H.-K. Chung and B. A. Hammel and L. F. Berzak Hopkins and L. C. Jarrott and S. F. Khan and B. Lahmann and R. Nora and M. J. Rosenberg and A. Pak and S. P. Regan and H. A. Scott and H. Sio and B. K. Spears and C. R. Weber},
title = {Development of a krypton-doped gas symmetry capsule platform for x-ray spectroscopy of implosion cores on the NIF},
journal = {Review of Scientific Instruments},
volume = {87},
number = {11},
pages = {11E327},
year = {2016},
doi = {10.1063/1.4960753},
}

@article{shi10,
author={Yuejiang Shi and Fudi Wang and Baonian Wan and Manfred Bitter and Sang Gon Lee and Jungyo Bak and Kenneth Hill and Jia Fu and Yingying Li and Ang Ti and Bili Ling}, 
title={Imaging x-ray crystal spectrometer on EAST},
journal={Plasma Phys. Control. Fusion},
volume=52, 
pages=085014,
year= 2010,
}

@article{hill10,
author={K. W. Hill and M. Bitter and L. Delgado-Aparicio and D. Johnson and R. Feder and P. Beiersdorfer and J. Dunn and K. Morris and E. Wang and M. Reinke and Y. Podpaly and J. E. Rice and R. Barnsley and M. OMullane and S. G. Lee}, 
journal=rsi,
title={Development of a spatially resolving x-ray crystal spectrometer for measurement of ion-temperature ($T_i$) and rotation-velocity ($v$) profiles in ITER},
volume=81,
pages={10E322},
year=2010,
}

@article{beiersdorfer10,
author={P. Beiersdorfer and J. Clementson and J. Dunn and M. F. Gu and K. Morris and Y. Podpaly and E. Wang and M. Bitter and R. Feder and K.W. Hill and D. Johnson and R. Barnsley},
title={he ITER core imaging x-ray spectrometer},
journal={J. Phys. B: At. Mol. Opt. Phys.},
volume=43,
pages=144008,
year=2010,
}

@article{Maria_PRE_2016,
  title = {Indications of flow near maximum compression in layered deuterium-tritium implosions at the National Ignition Facility},
  author = {Gatu Johnson, M. and Knauer, J. P. and Cerjan, C. J. and Eckart, M. J. and Grim, G. P. and Hartouni, E. P. and Hatarik, R. and Kilkenny, J. D. and Munro, D. H. and Sayre, D. B. and Spears, B. K. and Bionta, R. M. and Bond, E. J. and Caggiano, J. A. and Callahan, D. and Casey, D. T. and D\"oppner, T. and Frenje, J. A. and Glebov, V. Yu. and Hurricane, O. and Kritcher, A. and LePape, S. and Ma, T. and Mackinnon, A. and Meezan, N. and Patel, P. and Petrasso, R. D. and Ralph, J. E. and Springer, P. T. and Yeamans, C. B.},
  journal = {Phys. Rev. E},
  volume = {94},
  issue = {2},
  pages = {021202},
  numpages = {5},
  year = {2016},
  month = {Aug},
  publisher = {American Physical Society},
  doi = {10.1103/PhysRevE.94.021202},
  url = {https://link.aps.org/doi/10.1103/PhysRevE.94.021202}
}

@article{Spears_PoP_2015,
author = {Brian K. Spears and David H. Munro and Scott Sepke and Joseph Caggiano and Daniel Clark and Robert Hatarik and Andrea Kritcher and Daniel Sayre and Charles Yeamans and James Knauer and Terry Hilsabeck and Joe Kilkenny},
title = {Three-dimensional simulations of National Ignition Facility implosions: Insight into experimental observables},
journal = {Physics of Plasmas},
volume = {22},
number = {5},
pages = {056317},
year = {2015},
doi = {10.1063/1.4920957},
}

@article{Glebov_RSI_2010,
author = {V. Yu. Glebov and T. C. Sangster and C. Stoeckl and J. P. Knauer and W. Theobald and K. L. Marshall and M. J. {Shoup III} and T. Buczek and M. Cruz and T. Duffy and M. Romanofsky and M. Fox and A. Pruyne and M. J. Moran and R. A. Lerche and J. McNaney and J. D. Kilkenny and M. J. Eckart and D. Schneider and D. Munro and W. Stoeffl and R. Zacharias and J. J. Haslam and T. Clancy and M. Yeoman and D. Warwas and C. J. Horsfield and J.-L. Bourgade and O. Landoas and L. Disdier and G. A. Chandler and R. J. Leeper},
title = {The National Ignition Facility neutron time-of-flight system and its initial performance (invited)},
journal = {Review of Scientific Instruments},
volume = {81},
number = {10},
pages = {10D325},
year = {2010},
doi = {10.1063/1.3492351},
}

@article{lindl1995,
author = {John Lindl},
collaboration = {},
title = {Development of the indirect-drive approach to inertial confinement fusion and the target physics basis for ignition and gain},
publisher = {AIP},
year = {1995},
journal = {Physics of Plasmas},
volume = {2},
number = {11},
pages = {3933-4024},
url = {http://link.aip.org/link/?PHP/2/3933/1},
doi = {10.1063/1.871025}
}

@article{Lindl2004,
   author = "Lindl, John D. and Amendt, Peter and Berger, Richard L. and Glendinning, S. Gail and Glenzer, Siegfried H. and Haan, Steven W. and Kauffman, Robert L. and Landen, Otto L. and Suter, Laurence J.",
   title = "The physics basis for ignition using indirect-drive targets on the \mbox{N}ational \mbox{I}gnition \mbox{F}acility",
   journal = "Physics of Plasmas (1994-present)",
   year = "2004",
   volume = "11",
   number = "2", 
   pages = "339-491",
   url = "http://scitation.aip.org/content/aip/journal/pop/11/2/10.1063/1.1578638",
   doi = "http://dx.doi.org/10.1063/1.1578638" 
}

@article{Ingalls_JAP_1980,
author = {R. Ingalls and E. D. Crozier and J. E. Whitmore and A. J. Seary and J. M. Tranquada},
title = {Extended x-ray absorption fine structure of NaBr and Ge at high pressure},
journal = {Journal of Applied Physics},
volume = {51},
number = {6},
pages = {3158-3163},
year = {1980},
doi = {10.1063/1.328064},
}

@article{Bhalerao_JAP_2010,
author = {Gopalkrishna M. Bhalerao and Alain Polian and Michel Gauthier and Jean-Paul Iti{\'e} and Fran{\c c}ois Baudelet and Tapas Ganguli and Sudip K. Deb and Javed Mazher and Olivier Pag{\`e}s and Franciszek. Firszt and Wojciech Paszkowicz},
title = {High pressure x-ray diffraction and extended x-ray absorption fine structure studies on ternary alloy Zn1-xBexSe},
journal = {Journal of Applied Physics},
volume = {108},
number = {8},
pages = {083533},
year = {2010},
doi = {10.1063/1.3493850},
}

@Article{SanchezdelRio,
     author    = "S\'{a}nchez del Rio, M. and Dejus, R. J.",
     title     = "{XOP} 2.1---A New Version of the X-ray Optics Software Toolkit",
     journal = "AIP Conf. Proc.",
     year = "2004",
     volume = "705",
     pages = "784",
}

@Article{Stepanov,
     author    = "Stepanov, S.",
     title     = "X-ray server: an online resource for simulations of X-ray diffraction and scattering",
     journal = "Adv. Comp. Methods for X-ray and Neutron Optics",
     year = "2004",
     volume = "5536",
     pages = "16--26",
}

@article{Lu14,
   author = "Lu, J. and Bitter, M. and Hill, K. W. and Delgado-Aparicio, L. F. and Efthimion, P. C. and Pablant, N. A. and Beiersdorfer, P. and Caughey, T. A. and Brunner, J.",
   title = "X-ray tests of a two-dimensional stigmatic imaging scheme with variable magnifications",
   journal = "Review of Scientific Instruments",
   year = "2014",
   volume = "85",
   number = "11", 
   pages = "-", 
}

@article{Hill_RSI_2016,
author = {K. W. Hill and M. Bitter and L. Delgado-Aparicio and P. C. Efthimion and R. Ellis and L. Gao and J. Maddox and N. A. Pablant and M. B. Schneider and H. Chen and S. Ayers and R. L. Kauffman and A. G. MacPhee and P. Beiersdorfer and R. Bettencourt and T. Ma and R. C. Nora and H. A. Scott and D. B. Thorn and J. D. Kilkenny and D. Nelson and M. {Shoup III} and Y. Maron},
title = {Development of a high resolution x-ray spectrometer for the National Ignition Facility (NIF)},
journal = {Review of Scientific Instruments},
volume = {87},
number = {11},
pages = {11E344},
year = {2016},
doi = {10.1063/1.4962053},
}

@article{Hill_IAEA_2016,
 author={K. W. Hill and M. Bitter and P. C. Efthimion and R. Ellis and L. Gao and M. B. Schneider and H. Chen and S. Ayers and M. A. Barrios and P. Beiersdorfer and P. M. Bell and R. Bettencourt and D. K. Bradley and D. Casey and M. J. Edwards and B. A. Hammel and M. C. Hermann and W. W. Hsing and O. S. Jones and R. L. Kauffman and O. L. Landen and D. A. Liedahl and T. Ma and A. G. MacPhee and J. D. Moody and R. C. Nora and P. Patel and H. A. Scott and V. A. Smalyuk and B. K. Spears and D. B. THorn and J. D. Kilkenny and S. P. Regan and D. Nelson and R. Jungquist and M. {Shoup III} and Y. Maron and M. Sanchez del Rio},
 title={Adapting high resolution x-ray spectroscopy from MFE to temperature and density measurements in ICF},
 journal={26th IAEA Fusion Energy Conference, Kyoto, Japan, October, 2016},
}

@article{Hall_1984,
  author={T A Hall},
  title={A focusing X-ray crystal spectrograph},
  journal={Journal of Physics E: Scientific Instruments},
  volume={17},
  number={2},
  pages={110},
  url={http://stacks.iop.org/0022-3735/17/i=2/a=007},
  year={1984}
}

@article{Ma_RSI_2012,
author = {T. Ma and N. Izumi and R. Tommasini and D. K. Bradley and P. Bell and C. J. Cerjan and S. Dixit and T. D{\"o}ppner and O. Jones and J. L. Kline and G. Kyrala and O. L. Landen and S. LePape and A. J. Mackinnon and H.-S. Park and P. K. Patel and R. R. Prasad and J. Ralph and S. P. Regan and V. A. Smalyuk and P. T. Springer and L. Suter and R. P. J. Town and S. V. Weber and S. H. Glenzer},
title = {Imaging of high-energy x-ray emission from cryogenic thermonuclear fuel implosions on the NIF},
journal = {Review of Scientific Instruments},
volume = {83},
number = {10},
pages = {10E115},
year = {2012},
doi = {10.1063/1.4733313},
}

\end{document}